%% file: ms_pdf.tex
\documentclass[12pt,letterpaper,preprint]{aastex}
\usepackage{natbib}

\newcommand{\myemail}{p.lacerda@qub.ac.uk}
\newcommand{\tna} {\,\tablenotemark{\it a}}
\newcommand{\tnb} {\,\tablenotemark{\it b}}
\newcommand{\tnc} {\,\tablenotemark{\it c}}
\newcommand{\tnd} {\,\tablenotemark{\it d}}
\newcommand{\tne} {\,\tablenotemark{\it e}}
\newcommand{\tnf} {\,\tablenotemark{\it f}}
\newcommand{\tng} {\,\tablenotemark{\it g}}

\slugcomment{Accepted, 18 July 2011} \shorttitle{A Change in the Lightcurve of
Contact Binary KBO 2001$\,$QG$_{298}$} \shortauthors{Pedro Lacerda}

\begin{document}

\def\FigOne{
   \begin{figure}
   \centering
      \includegraphics[width=1.00\textwidth]{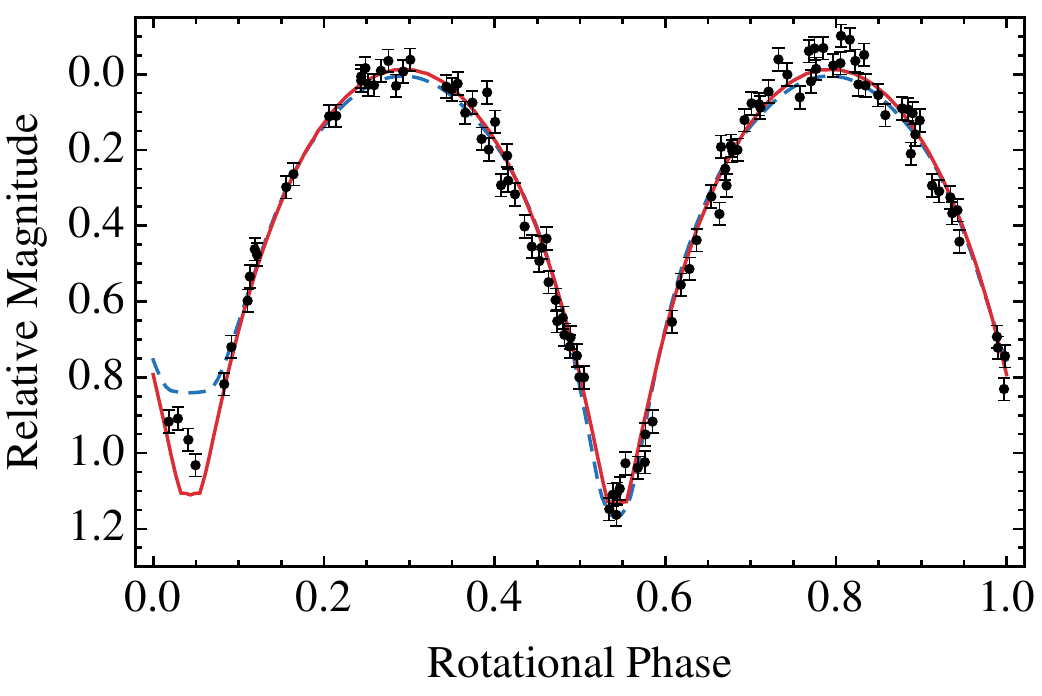}

   \caption[FigOne] {KBO 2001$\,$QG$_{298}$ lightcurve data (black points)
   obtained in 2003 by \citet{2004AJ....127.3023S}. Fits based on Roche
   binaries are shown as solid lines \citep{2007AJ....133.1393L}. Blue (dashed)
   line assumes an icy-type surface and red (solid) line assumes a lunar-type
   surface. The model binaries are rendered in Fig.\ \ref{Fig.Four}.} 

   \label{Fig.One}
   \end{figure}
}

\def\FigTwo{
   \begin{figure}
   \centering
      \includegraphics[width=1.00\textwidth]{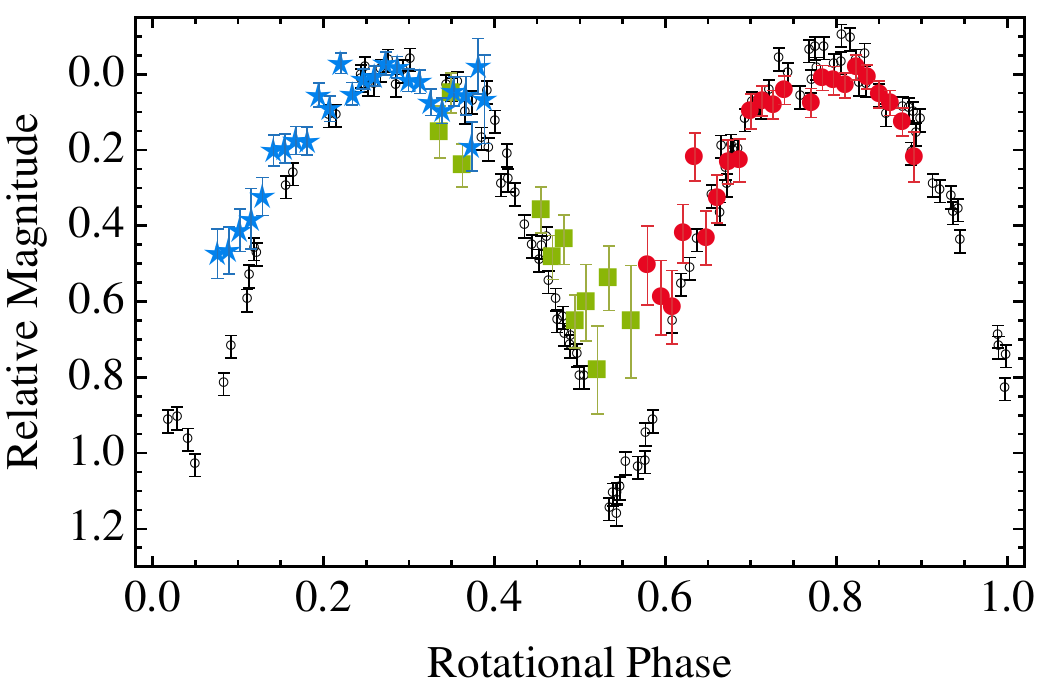}

   \caption[FigTwo] {Lightcurves of KBO 2001$\,$QG$_{298}$ in 2003  \citep[open
   circles;][]{2004AJ....127.3023S} and 2010 (filled symbols; this work).  Red
   circles, green squares and blue stars indicate data from 2010 August 15, 16
   and 17, respectively. The photometric range has decreased from $\Delta
   m_{2003}=1.14\pm0.04$ mag in 2003 to $\Delta m_{2010}=0.7\pm0.1$ mag in
   2010.  The 2003 and 2010 lightcurves appear in phase.} 

   \label{Fig.Two}
   \end{figure}
}

\def\FigThree{
   \begin{figure}
   \centering
      \includegraphics[width=1.00\textwidth]{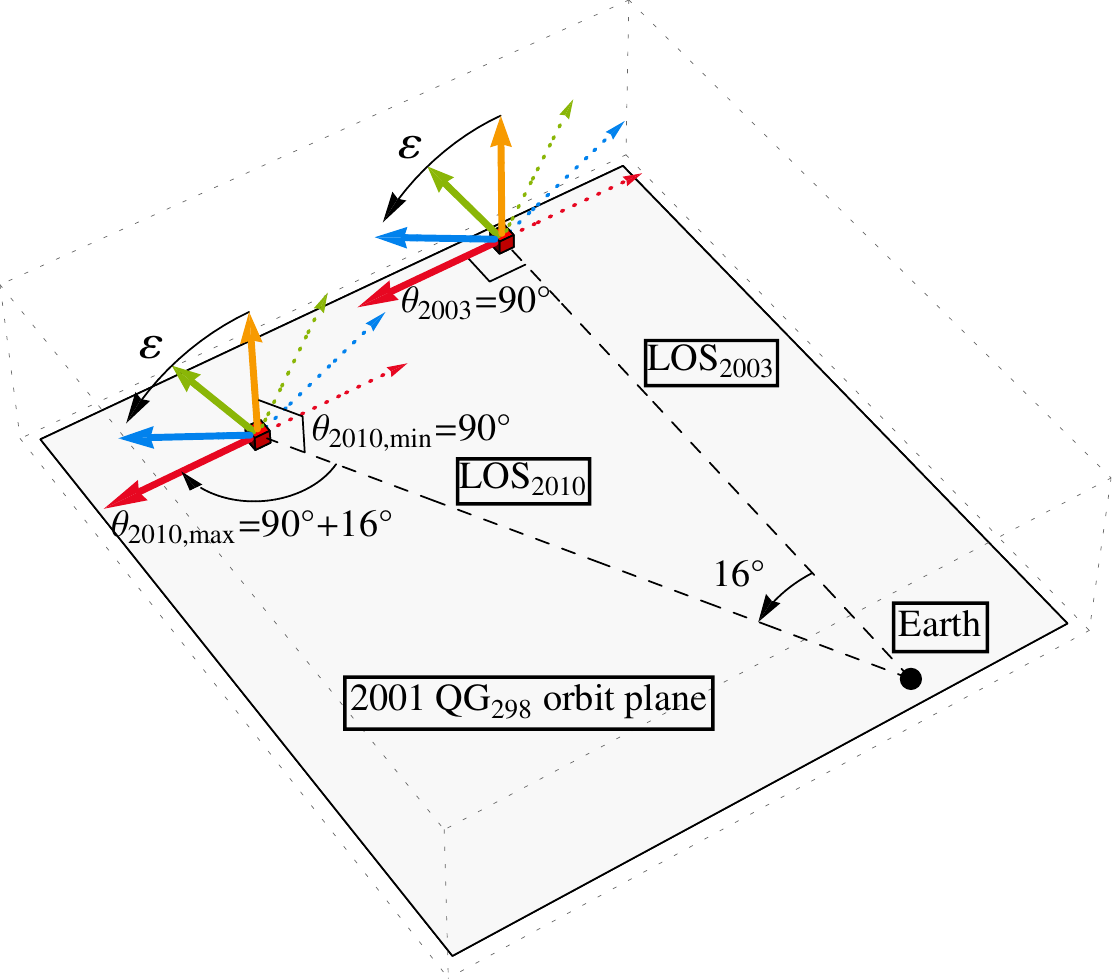}

   \caption[FigThree] {Geometric circumstances and relevant angles used to
   describe the lightcurve observations of 2001$\,$QG$_{298}$ in 2003 and 2010.
   The orbit plane of 2001$\,$QG$_{298}$ is shaded and labelled. The location
   of the Earth (black dot) and the lines of sight (dashed lines
   ``LOS$_{2003}$'' and ``LOS$_{2010}$'') towards 2001$\,$QG$_{298}$ are also
   indicated. The positions of 2001$\,$QG$_{298}$ in 2003 and 2010 are marked
   by red boxes. A few possible spin pole orientations (obliquities
   $\varepsilon=0\degr,30\degr,60\degr,90\degr$ measured from the normal to the
   orbit plane, as indicated) for 2001$\,$QG$_{298}$ are illustrated as thick
   solid arrows emerging from the object. Thinner dashed lines correspond to
   symmetric pole orientations (w.r.t.  $\varepsilon=0\degr$) which produce
   similar lightcurves to their counterparts and cannot be observationally
   distinguished. Under the assumption that the 2003 aspect angle
   $\theta_{2003}=90\degr$, the 2010 aspect angle lies in the range
   $90\degr\leq\theta_{2010}\leq90\degr+16\degr$ (or equivalently
   $90\degr-16\degr\leq\theta_{2010}\leq90\degr$) depending on the obliquity.}

   \label{Fig.Three}
   \end{figure}
}

\def\FigFour{
   \begin{figure}
   \centering
      \includegraphics[width=1.00\textwidth]{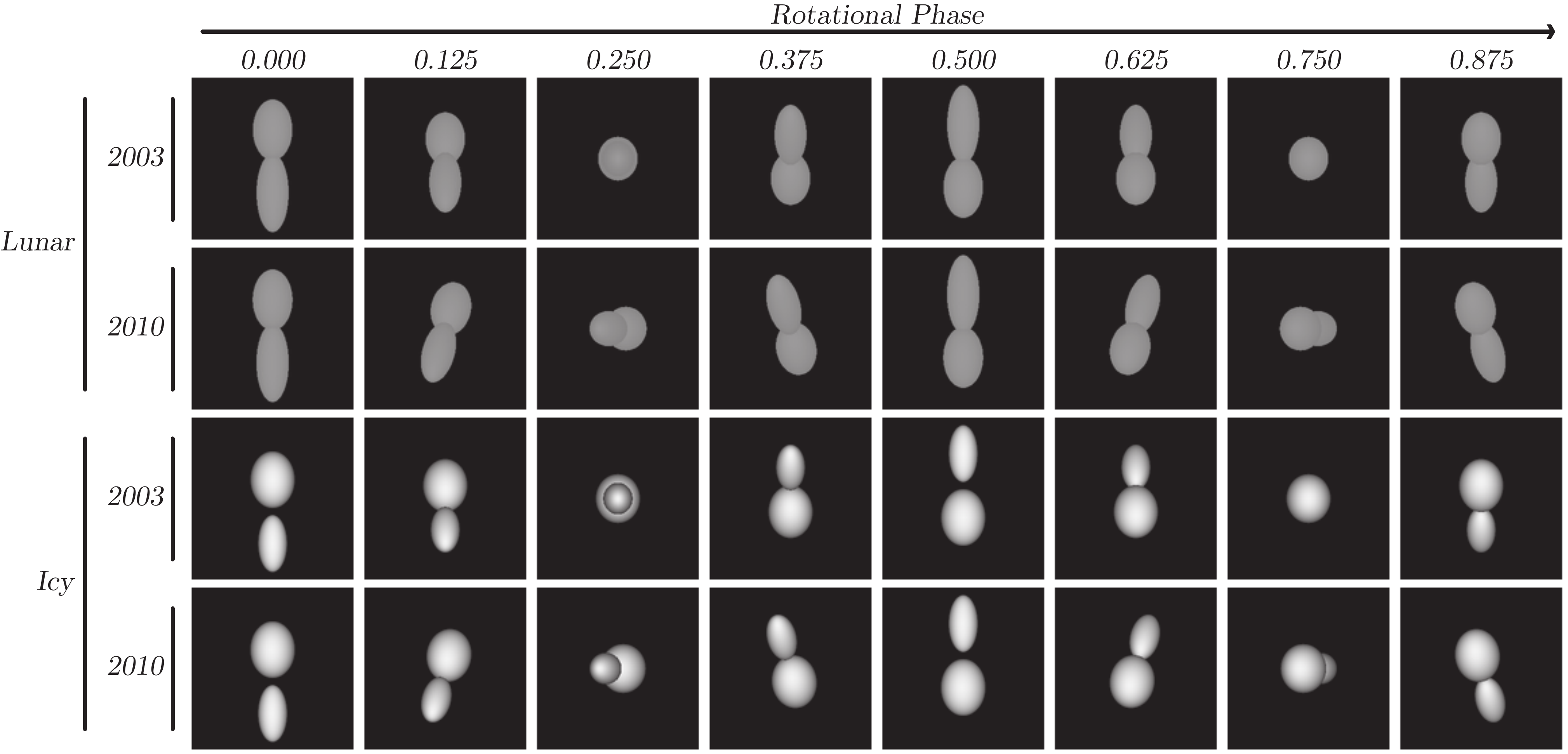}

   \caption[FigFour] {Renderings of the Roche binary models of
   2001$\,$QG$_{298}$ \citep{2007AJ....133.1393L} assuming lunar-type (top two
   rows) and icy-type (bottom two rows) surface scattering. The models are
   rendered as seen from Earth in 2003 (rows 1 and 3 from top) and in 2010
   (rows 2 and 4 from top) assuming that 2001$\,$QG$_{298}$ has obliquity
   $\varepsilon=90\degr$. Successive rotational phases are displayed from left
   to right to simulate rotation and to show maximum and minimum cross-section
   configurations.} 

   \label{Fig.Four}
   \end{figure}
}

\def\FigFive{
   \begin{figure}
   \centering
      \includegraphics[width=0.80\textwidth]{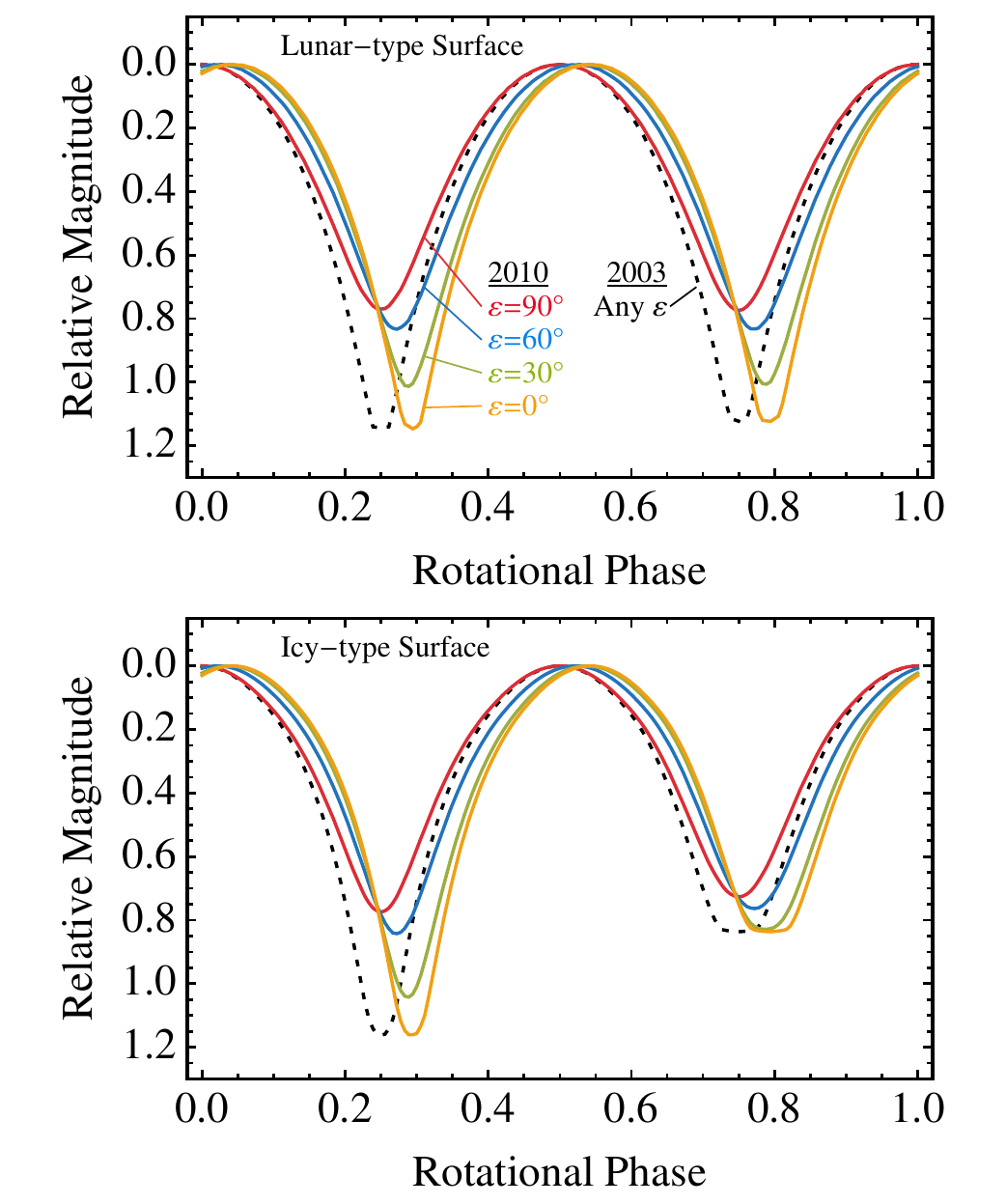}

   \caption[FigFive] {2001$\,$QG$_{298}$ lightcurve model predictions for 2010
   (solid, colored lines) based on the 2003 models (dashed lines). The
   predicted lightcurve range and phase depend on the assumed obliquity
   $\varepsilon$. Lunar-type (icy-type) scattering models are shown in the top
   (bottom) panel. } 

   \label{Fig.Five}
   \end{figure}
}

\def\FigSix{
   \begin{figure}
   \centering
      \includegraphics[width=1.00\textwidth]{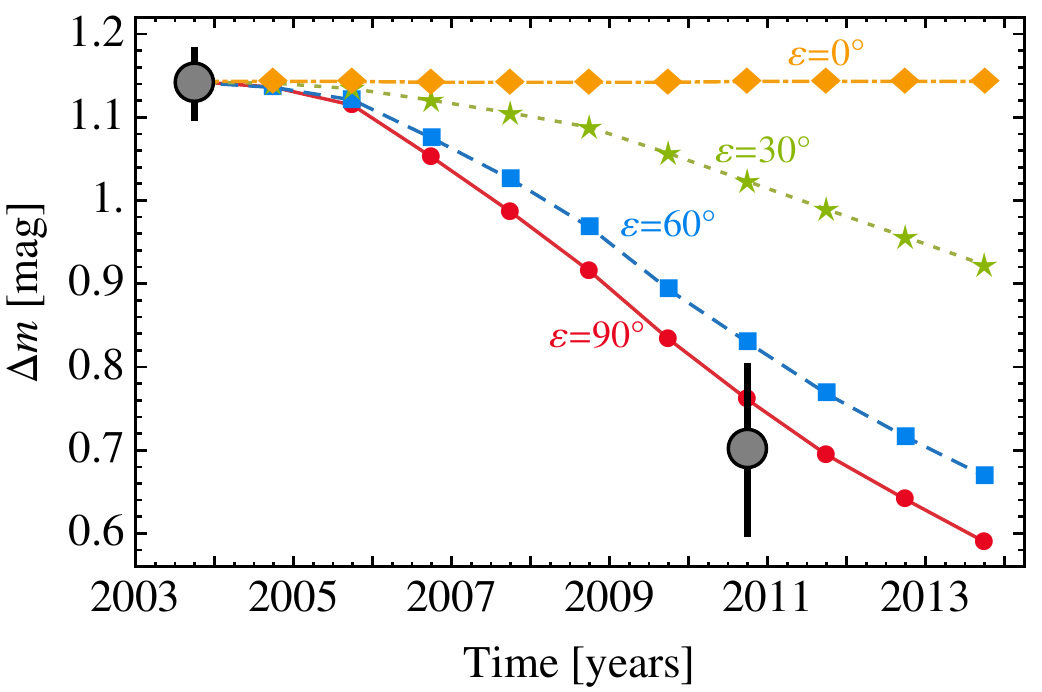}

   \caption[FigSeven] {Predicted photometric range for 2001$\,$QG$_{298}$ as a
   function of time and obliquity. Measured 2003 and 2010 photometric ranges
   are shown as large gray points with 1-$\sigma$ error bars. Table
   \ref{Table.Four} lists the $\Delta m$ values as a function of obliquity at
   the time of the 2010 measurement. } 

   \label{Fig.Six}
   \end{figure}
}

\def\FigSeven{
   \begin{figure}
   \centering
      \includegraphics[width=1.00\textwidth]{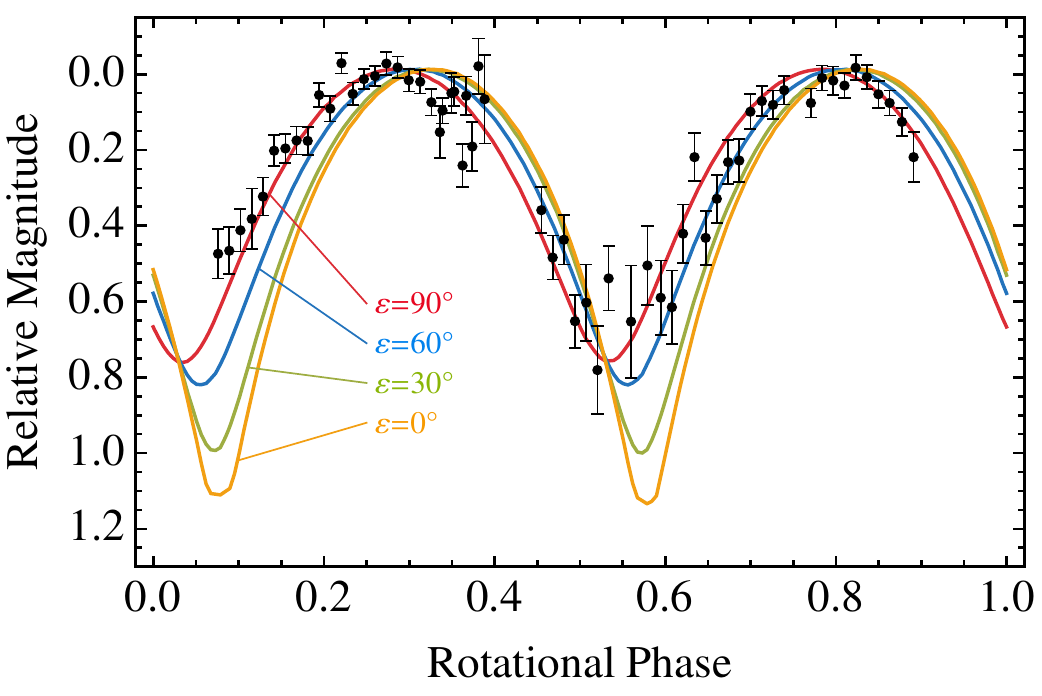}

   \caption[FigSix] {Comparison between the 2010 lightcurve data (black dots
   with error bars) and the model predictions shown in Fig.\ \ref{Fig.Five}.
   Only the lunar-type scattering models are plotted. Different lines assume
   obliquities $\varepsilon=0\degr,30\degr,60\degr,90\degr$.  Respective
   reduced $\chi^2$ values for the fits are
   $\chi^2_\mathrm{red}=18.8,12.3,4.4,1.7$, favoring the solution
   $\varepsilon=90\degr$.} 

   \label{Fig.Seven}
   \end{figure}
}

\def\FigEight{
   \begin{figure}
   \centering
      \includegraphics[width=1.00\textwidth]{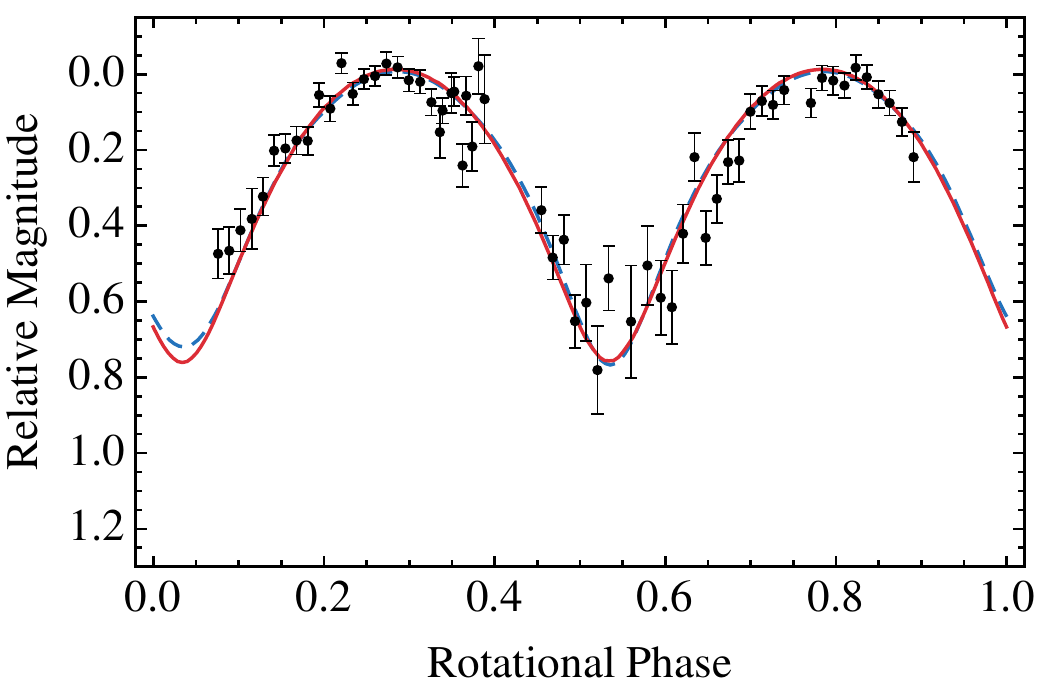}

   \caption[FigEight] {Negligible difference between the lunar- (red, solid)
   and icy-type (blue, dashed) scattering model lightcurves.} 

   \label{Fig.Eight}
   \end{figure}
}

\def\FigNine{
   \begin{figure}
   \centering
      \includegraphics[width=1.00\textwidth]{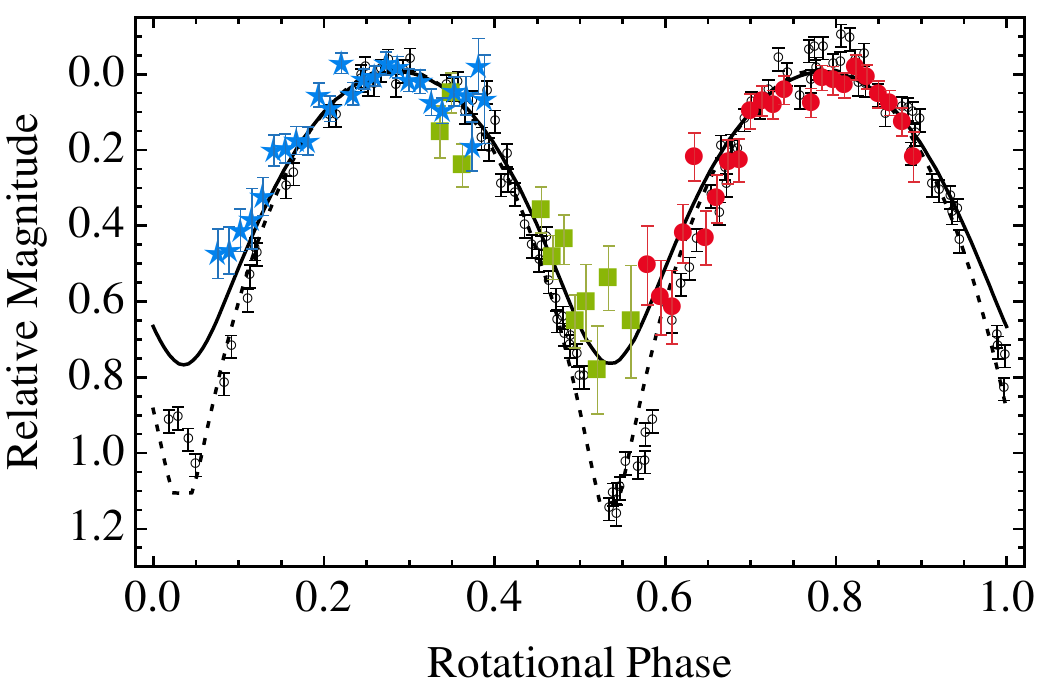}

   \caption[FigNine] {Lightcurves from 2003 and 2010 (as shown in Fig.\
   \ref{Fig.Two}) simultaneously fit by the model in Fig.\ \ref{Fig.Four} (top
   two rows). The model assumes obliquity $\varepsilon=90\degr$.} 

   \label{Fig.Nine}
   \end{figure}
}

\def\FigTen{
   \begin{figure}
   \centering
      \includegraphics[width=0.75\textwidth]{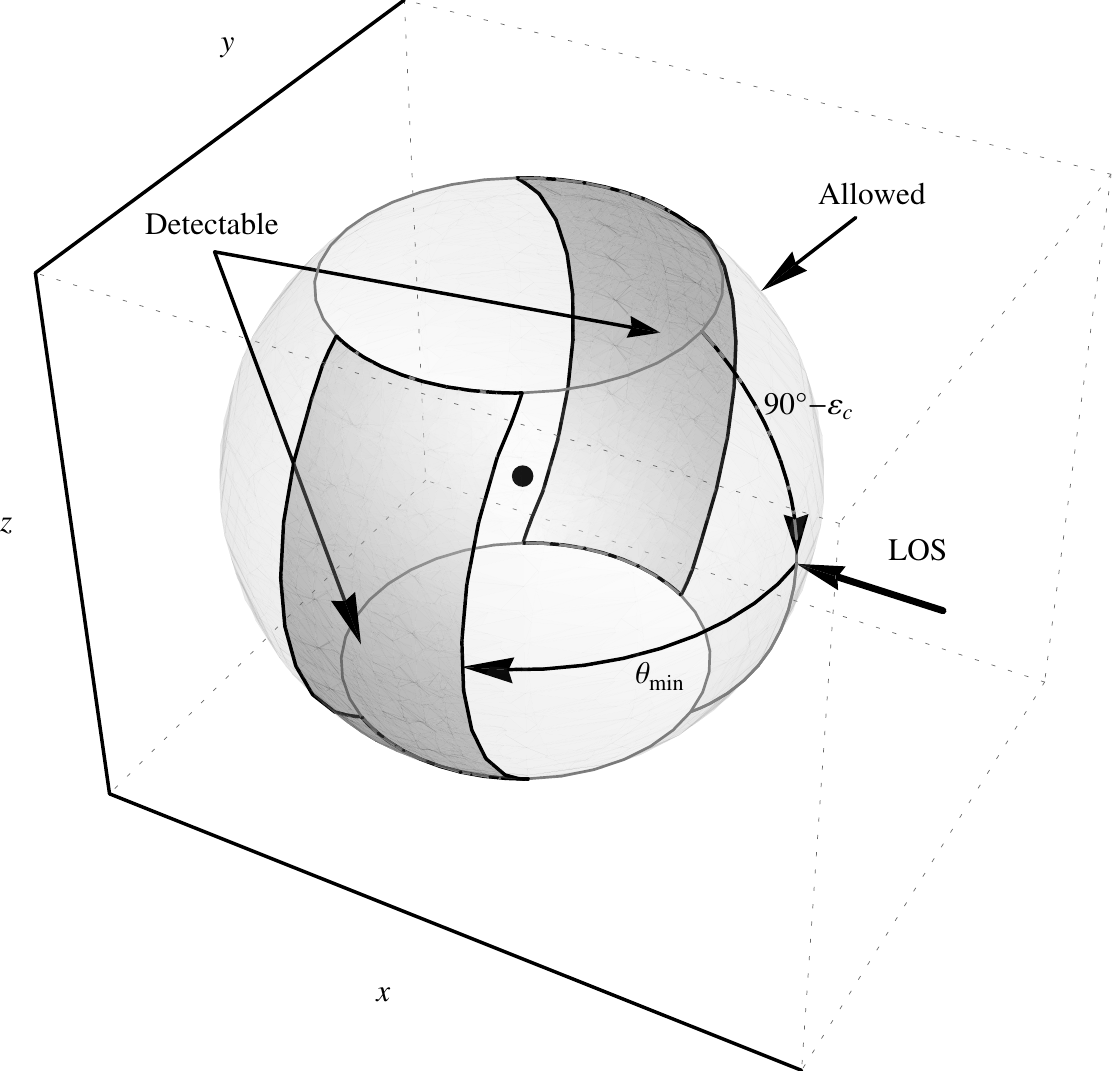}

   \caption[FigTen] {Solid angles used to calculate the probability that a
   contact binary with obliquity in the range
   $\varepsilon_c<\varepsilon<180\degr-\varepsilon_c$ ($\varepsilon_c=40\degr$)
   will be detected from its high photometric variability. The line of sight is
   indicated by a thick, black arrow labelled ``LOS''. Solid angle regions of
   interest are marked on the surface of a sphere surrounding the contact
   binary (marked by a black dot). The solid angle region where spin poles are
   allowed with uniform probability is shaded in light gray. That region is
   defined as the full sphere minus polar caps extending an angle
   $\varepsilon_c$ from the either pole. Darker gray regions mark spin pole
   orientations that are detectable from Earth. Those regions lie more than
   $\theta_\mathrm{min}$ from the LOS. The ratio of these “detectable” and
   “allowed” solid angles equals the detection probability.} 

   \label{Fig.Ten}
   \end{figure}
}

\title{A Change in the Lightcurve of Kuiper Belt Contact Binary (139775)
2001$\,$QG$_{298}$}

\author{Pedro Lacerda}

\affil{Astrophysics Research Centre, Queen's University Belfast, Belfast BT7
1NN, UK}

\email{\myemail}

\begin{abstract}

New observations show that the lightcurve of Kuiper belt contact binary
(139775) 2001$\,$QG$_{298}$ has changed substantially since the first
observations in 2003. The 2010 lightcurve has a peak-to-peak photometric of
range $\Delta m_{2010}=0.7\pm0.1$ mag, significantly lower than in 2003,
$\Delta m_{2003}=1.14\pm0.04$ mag. This change is most simply interpreted if
2001$\,$QG$_{298}$ has an obliquity near $90\degr$. The observed decrease in
$\Delta m$ is caused by a change in viewing geometry, from equator-on in 2003
to  nearly 16$\degr$ (the orbital angular distance covered by the object
between the observations) off the equator in 2010. The 2003 and 2010
lightcurves have the same rotation period and appear in phase when shifted by
an integer number of full rotations, also consistent with high obliquity. Based
on the new 2010 lightcurve data, we find that 2001$\,$QG$_{298}$ has an
obliquity $\varepsilon=90\degr\pm30\degr$.  Current estimates of the intrinsic
fraction of contact binaries in the Kuiper belt are debiased assuming that
these objects have randomly oriented spins. If, as 2001$\,$QG$_{298}$, KBO
contact binaries tend to have large obliquities, a larger correction is
required. As a result, the abundance of contact binaries may be larger than
previously believed.

\end{abstract}


\section{Introduction}

Located approximately between 30 and 50 AU from the Sun, the Kuiper belt
represents the outer frontier of the currently observable solar system.  The
belt is estimated to hold roughly forty thousand objects larger than 100 km in
diameter
\citep{1995AJ....109.1867Jewitt,2001AJ....122..457T,2004AJ....128.1364B,2008AJ....136...83Fuentes,2008Icar..195..827Fraser}.
More than one thousand objects have been detected but only six hundred have
multi-opposition observations and well-determined orbits. The Kuiper belt
objects (KBOs) are remnants of outer solar system planetesimals and their study
may shed light on the complex process that led to the formation of planets.
Our understanding of the physical properties of KBOs remains rudimentary as
most are too faint for detailed investigation. 

Many KBOs are binary.  Among the dynamically cold, ``classical'' population
(low eccentricity and inclination KBOs located between the 3:2 and the 2:1
mean-motion resonances with Neptune, at 39.4 and 47.8 AU) the binary frequency
is inferred to be $\sim$20\% whereas in other dynamical subsets the fraction is
lower, at 5 to 10\% \citep{2006AJ....131.1142Stephens,2008Icar..194..758Noll}.
Binaries probably formed early on \citep{2002Icar..160..212W}, before the
Kuiper belt lost more than 99\% of its original mass
\citep{1998AJ....115.2136K,2009MNRAS.399..385Booth}, because the current-day
number density of KBOs is too low for frequent pair encounters. It is also
possible that planetesimals formed in clusters of two or more bodies and that
the binaries we see today are direct remnants of that process
\citep{2010AJ....140..785Nesvorny}. Thus, binaries are particularly useful
probes of the conditions under which planetesimals formed. For instance, the
symmetry in surface properties of KBO pairs argues for a chemically homogeneous
formation environment \citep{2009Icar..200..292Benecchi}. But the current
binary abundance also bears record of the effects of collisional and dynamical
processes that have eroded the original population \citep{2004Icar..168..409P}.
It is highly desirable to find ways to isolate the effects of formation,
evolution, and erosion.  The relative frequency of binaries within different
regions of the Kuiper belt, and their distributions of orbital properties and
relative mass of the components are quantities that should be sought to help
clarify the formation and evolution of binaries
\citep{2008ApJ...686..741Schlichting,2011ApJ...730..132Murray-Clay,2011AJ....141..159Nesvorny}. 

Most known binaries are resolved \citep[physical separations $>$1500
km;][]{2002AJ....124.3424N,2006AJ....131.1142Stephens,2006ApJ...643..556Stansberry,2011arXiv1103.2751Grundy}.
The exception is the $500\,\mathrm{km}\times175\,\mathrm{km}$ contact binary
(139775) 2001$\,$QG$_{298}$, identified from analysis of its rotational
lightcurve \cite[Figure \ref{Fig.One};][hereafter SJ04]{2004AJ....127.3023S}.
Using time-resolved measurements taken mainly during 2003, the authors found
that 2001$\,$QG$_{298}$ displayed extreme photometric variability ($\Delta
m\sim1.2$ mag) and a relatively slow rotation period ($P\sim13.77$ hr). The
large photometric range, caused by the eclipsing nature of the binary, exceeds
the $\Delta m=0.75$ mag produced by two spheres in contact (axis ratio
$b/a=0.5$) because the contact binary components are tidally elongated along
the line connecting the centers
\citep{1980Icar...44..807W,1984A&A...140..265L}. The large $\Delta m$ found for
2001$\,$QG$_{298}$ also implies that the system was observed almost
equatorially in 2003. Models of 2001$\,$QG$_{298}$'s lightcurve based on
figures of hydrostatic equilibrium lend strong support to these assertions
\citep{2004PASJ...56.1099T,2007AJ....133.1393L,2010ApJ...719.1602Gnat} and
allow the bulk density of 2001$\,$QG$_{298}$ to be estimated at
$\rho=0.59_{-0.05}^{+0.14}$ g$\,$cm$^{-3}$. This surprisingly low density
implies that 2001$\,$QG$_{298}$ is mostly icy in composition and is
significantly porous.

The example of 2001$\,$QG$_{298}$ suggests that between 10\% and 20\% of KBOs
could be contact binaries, depending on surface scattering properties
\citep[SJ04;][]{2007AJ....133.1393L}. This high fraction is consistent with the
observation that the frequency of resolved binaries increases dramatically with
decreasing orbital separation \citep{2006ApJ...643L..57Kern}.  However, despite
their high abundance, contact binary KBOs remain elusive. The reason is that
these objects are unresolved and can only be identified if they happen to be
nearly equator-on. The consequence is that as many as 85\% of contact binaries
may go unnoticed due to unfavorable observing geometry. High phase angle
observations that have been shown to increase the chance of detecting contact
binaries \citep{2008ApJ...672L..57Lacerda} can not be applied to the distant
KBOs as their phase angles remain below 2\degr\ when observed from Earth.
Interestingly, the Jupiter Trojans also contain a fairly large fraction of
contact binaries, comparable to that found for KBOs. In addition to the
well-known case of (624) Hektor
\citep{1978Icar...36..353H,1994A&A...281..269Detal,2001Icar..153..348Cruikshank,2007AJ....133.1393L},
two other contact binaries have been identified in a survey of 114 Trojans
\citep{2007AJ....134.1133Mann}. When corrected by the probability of observing
these objects close to equator-on, \citet{2007AJ....134.1133Mann} find an
intrinsic contact-binary fraction of at least 6\% to 10\%. The apparently high
abundance of contact binaries and the prospect of measuring their densities
makes these primitive objects particularly interesting targets of study.

QG$_{298}$ has travelled a significant angular distance ($\sim16\degr$) in its
heliocentric orbit since the first observations in 2003. In this paper we
compare new 2010 lightcurve data with the SJ04 observations and make inferences
on the obliquity of 2001$\,$QG$_{298}$. In Section 2 we describe our
observations, in Section 3 we report our findings on the lightcurve and
obliquity of 2001$\,$QG$_{298}$, in Section 4 we discuss the implications of
our findings for the abundance of contact binaries, and in Section 5 we present
our conclusions.

\section{Observations}

Visible-light observations were taken at the 2.5-m Isaac Newton Telescope
(INT) operating on the island of La Palma in the Canary Islands, Spain. The INT
was equipped with the Wide Field Camera (WFC), mounted at the f/3.29 prime
focus. The WFC consists of an array of 4 thinned EEV 4K$\times$2K
charge-coupled devices (CCDs). The CCDs have an image scale of 0.33 arcsec per
pixel and the full array covers approximately 34 arcmin squared on the sky.
Our observations made use of CCD \#4 which is the most central with respect to
the field of view.  Details of the observing times and conditions can be found
in Table\ \ref{Table.One}.

Our target, 2001$\,$QG$_{298}$, was observed through the Sloan-Gunn $r$ filter
(Isaac Newton Group filter \#214) which approximates the $r$ band in the Sloan
Digital Sky Survey (SDSS) photometric system. Consecutive 600 s exposures were
obtained in a consistent manner to measure variations in the brightness of
QG$_{298}$. The rate of motion of 2001$\,$QG$_{298}$ across the sky was $<$2.32
arcsec/hr (0.39 arcsec in 10 minutes) keeping it well within the seeing disc
during the observations. All frames were bias-subtracted and flat-fielded using
the median of a set of dithered exposures taken during twilight. In the reduced
images, the brightness variations of 2001$\,$QG$_{298}$ were measured
differentially against 8 bright, not saturated, field stars to offer protection
from fluctuations in atmospheric transmission and seeing.  The mutual relative
photometry of the comparison stars is stable to better than $\pm$0.01 mag
throughout all the observations.  The field containing 2001$\,$QG$_{298}$
($\mathrm{RA}=0\fh838$, $\mathrm{dec}=1\fdg964$) at the time of our
observations is included in the SDSS photometric catalog. This allowed us to
absolutely calibrate our observations directly against SDSS magnitudes of 5 of
the 8 comparison stars (see Table\ \ref{Table.Two}) chosen for their spatial
distribution and exceptional photometric stability.

\section{Results}

\subsection{The lightcurve of 2001$\,$QG$_{298}$}

We used our photometric measurements of 2001$\,$QG$_{298}$ to construct a
lightcurve (magnitudes versus time) for this object.  Figure\ \ref{Fig.Two}
shows how our 2010 lightcurve and the SJ04 lightcurve compare. Overall, the
lightcurves have the same period and appear aligned in rotational phase. The
2010 photometric range is smaller by about 0.5 magnitudes.

The apparent alignment in rotational phase between the 2003 and 2010
lightcurves is real. All times were measured relative to 2010 Aug 13 at 0h UT
(zero rotational phase) and then phased to the period $P=13.7744\pm0.0002$ hr
(SJ04). The maximum light-travel time difference between any two measurements
included in the 2003 and 2010 lightcurves is less than 1.5 minutes (0.002
rotations) so no delay correction was required. More importantly, the relative
uncertainty in the rotational period of 2001$\,$QG$_{298}$ translates to an
uncertainty of 0.06 rotations between the 2003 and 2010 data.  The latter
implies that the phase alignment in Figure \ref{Fig.Two} is accurate to
$\pm0.06$ in rotational phase.

The vertical axis in Figure \ref{Fig.Two} indicates magnitudes relative to the
maximum apparent brightness of 2001$\,$QG$_{298}$. The vertical alignment of
the 2003 and 2010 data is artificially set to facilitate comparison of the
photometric ranges in 2003 and 2010 ($\Delta m_{2003}$ and $\Delta m_{2010}$).
In our data, we find that at maximum brightness 2001$\,$QG$_{298}$ has
magnitude $m_r=21.63\pm0.02$ mag. This measurement is absolutely calibrated
using SDSS catalog stars as described in Section 2 and the error is dominated
by uncertainties in the $r$ magnitudes of the calibration stars (Table\
\ref{Table.Two}). SJ04 found that at maximum brightness 2001$\,$QG$_{298}$ has
an absolute magnitude $m_R(1,1,0)=6.28\pm0.02$ mag. The expected apparent magnitude at the time of our observations is thus

\[ m_R(R,\Delta,\alpha)=m_R(1,1,0)+5\log R \Delta+\beta \alpha=21.46\pm0.04 \
\mathrm{mag}, \]

\noindent where $R$, $\Delta$ and $\alpha$ are taken from Table\
\ref{Table.One}, and $\beta=0.15\pm0.01$ mag/\degr\ is the phase coefficient
for KBOs \citep{2002AJ....124.1757S}. To compare the expected $m_R$ magnitude
with our new $m_r$ measurement we used two approaches. Firstly, we converted
our SDSS $r$ magnitude to a Cousins $R$ magnitude using the transformation
equations in \citet{2002AJ....123.2121Smith} and the broadband colors of
2001$\,$QG$_{298}$ ($B-V=1.00\pm0.04$ mag and $V-R=0.60\pm0.02$ mag; see SJ04):
we obtain an equivalent $R$ magnitude $m_R=21.36\pm0.04$ mag.  Secondly, we
took the transmission profiles of the two filters (available from the Isaac
Newton Group online filter database) and convolved each with the spectrum of
2001$\,$QG$_{298}$ to find an integrated flux ratio 1.215. This corresponds to
magnitude difference $m_r-m_R=0.21$ mag between the filters. Subtracting that
difference from our $m_r$ measurement we obtain an equivalent $R$ magnitude
$m_R=21.42\pm0.04$ mag. We assumed the spectrum of 2001$\,$QG$_{298}$ to be
linear with a slope 22.6\% (1000\AA)$^{-1}$, as calculated from the object's
$B-V$ color. Although both approaches are uncertain, they agree at the
1-$\sigma$ level and fall within 0.1 mag of the expected apparent magnitude
based on the object's absolute magnitude and geometric circumstances.

\subsection{The obliquity of 2001$\,$QG$_{298}$}

Between the time of the SJ04 observations in 2003 and the time of our 2010
observations 2001$\,$QG$_{298}$ has travelled 16\degr\ along its heliocentric
orbit. A diagram of the geometric circumstances of the two sets of observations
is shown in Figure \ref{Fig.Three}. The change in observing geometry between
2003 and 2010 has predictable effects on the lightcurve of 2001$\,$QG$_{298}$.
Given the extreme variability displayed in 2003, 2001$\,$QG$_{298}$ was
probably observed equator-on, i.e.\ at aspect angle (measured between the
line of sight and the spin pole) $\theta_{2003}\sim90\degr$ (Figure
\ref{Fig.Three}).  This assertion is backed by detailed models of the 2003
lightcurve based on figures of hydrostatic equilibrium
\citep{2004PASJ...56.1099T,2007AJ....133.1393L,2010ApJ...719.1602Gnat}. Those
simulations consider close and contact binaries in which the components are
homogeneous and strengthless, and orbit each other in tidally locked, circular
orbits. Under these conditions, the binary components assume shapes that
balance gravitational (including mutual tidal) and rotational accelerations.
Furthermore, the exact shapes of the binary components, their mutual orbit
period ($\equiv$ spin period), and their density are all interdependent.  The
practical result is that the density can be inferred from the spin period and
the component shapes. The simulations search the space of hydrostatic
equilibrium binary solutions and orientations (aspect angles) for the
combination that best fits the observed lightcurve and then use the spin period
to extract the density. This technique has been tested on the observationally
well-characterized Jovian Trojan contact binary (624) Hektor
\citep{2007AJ....133.1393L}, using observations at multiple geometries along
its 12-yr orbit \citep{1969AJ.....74..796D}.

In the case of 2001$\,$QG$_{298}$, all attempts to simulate the SJ04 lightcurve
yield equivalent results and agree upon a nearly equator-on geometry, and a low
bulk density $\rho\sim0.6-0.7$ g/cm$^{-3}$
\citep{2004PASJ...56.1099T,2007AJ....133.1393L,2010ApJ...719.1602Gnat}. For our
purposes, we will use the models of 2001$\,$QG$_{298}$ found by
\citet[][hereafter LJ07]{2007AJ....133.1393L}; the SJ04 data and respective
LJ07 model lightcurves are shown in Figure \ref{Fig.One}. The binary models are
rendered in Figure \ref{Fig.Four}. LJ07 make use of the Roche approximation
which calculates the deformations on each component independently, assuming the
other component's mass is concentrated on a point
\citep{1963ApJ...138.1182C,1984A&A...140..265L}. In the Roche binary
approximation the binary is composed of two triaxial ellipsoids described by
their axis ratios, their mass ratio, and their orbital separation.  LJ07
include the effects of limb darkening when simulating the Roche binary
lightcurve. They consider two extreme light scattering laws: a lunar-type
(Lommel-Seeliger or backscatter reflection) law, appropriate for low albedo,
rocky surfaces, and an icy-type (Lambertian or diffuse reflection) law,
suitable for high albedo surfaces. The lunar law leads to negligible limb
darkening and so generates symmetric lightcurves for which both mutual eclipses
produce the same signature. The icy law causes considerable limb darkening and
so produces asymmetric lightcurves if the binary components have different
masses/sizes.  The SJ04 hints at a slight difference between the two lightcurve
minima (Figure \ref{Fig.One}), intermediate to the extreme lunar and icy cases,
so LJ07 identify two models, one for each scattering law, that bracket the data
(see Figure \ref{Fig.One}).  The model parameters are listed in Table
\ref{Table.Three}. 

We want to investigate how the lightcurve is predicted to change between 2003
and 2010 depending on the binary's obliquity, $\varepsilon$ (see Figure
\ref{Fig.Three}). The obliquity is the angle between the outer (heliocentric)
orbit and inner (mutual) orbit planes.  We shall refer to the normals to the
outer orbit and inner orbit planes of the binary as orbit pole and spin pole,
respectively.  Assuming that 2001$\,$QG$_{298}$ was observed exactly equator-on
in 2003, then the spin pole must lie on the plane perpendicular to the 2003
line of sight (aspect angle $\theta_{2003}=90\degr$) and the obliquity is
simply the sky-projected angle between spin pole and the normal to the
ecliptic.  [Note: The Earth is less than
$\left(1\,\mathrm{AU}/31\,\mathrm{AU}\right)\left(180\degr/\pi\right)\approx1.8\degr$
away from the orbit plane of 2001$\,$QG$_{298}$ as seen from the KBO, which is
negligible.] In this configuration, the lightcurve has the same shape
independent of the obliquity.  In 2010 the spin pole of 2001$\,$QG$_{298}$
makes an angle with the line of sight
$\theta_{2010}=\arccos\left(-\sin\varepsilon\,\sin\Delta\nu\right)$ where
$\Delta\nu=16\degr$ is the angle through which 2001$\,$QG$_{298}$ has moved
along its orbit since 2003. The 2010 aspect angle can take values in the range
$90\degr\leq\theta_{2010}\leq90\degr+16\degr$ depending on its
obliquity\footnote{Note that we are considering the case (shown as thick solid
vectors in Figure \ref{Fig.Three}) in which the pole vector moves away from the
observer between 2003 and 2010; the opposite case in which the pole moves
towards (and $90\degr-16\degr\leq\theta_{2010}\leq90\degr$) the observer (thin
dashed vectors in Figure \ref{Fig.Three}) produces a similar effect on the 2010
lightcurve.}. In other words, the equator of 2001$\,$QG$_{298}$ lies at most
16$\degr$ from the line of sight in 2010.

We took the two models in Table \ref{Table.Three} and rendered them as they
would appear from Earth in 2010 (Figure \ref{Fig.Four}) following the procedure
in LJ07. The resulting 2010 model lightcurves are shown in Figure
\ref{Fig.Five}.  As expected, at obliquity $\varepsilon=0\degr$ the lightcurve
displays no change in photometric range, $\Delta m_{2003}=\Delta m_{2010}$.
However, the lightcurve phase is shifted (delayed) by
$16\degr/360\degr\approx0.044$ as the object must rotate an extra 16\degr\ in
order to appear at the same rotational phase as seen from Earth.  As the
obliquity increases, the 2010 photometric range decreases, down to a minimum of
$\Delta m_{2010}\sim0.7$ mag for $\varepsilon=90\degr$. The rotational phase
shift also decreases as $\varepsilon$ increases and for $\varepsilon=90\degr$
the shift is zero and the 2003 and 2010 lightcurves appear in phase. As the
obliquity approaches $\varepsilon=90\degr$ the 2010 lightcurve looks nearly
identical for the lunar- and icy-type surfaces. This occurs because the minimum
cross-section configuration becomes less affected by limb darkening effects
(see Figure \ref{Fig.Four}).

Figure \ref{Fig.Six} shows how the observed 2003 and 2010 photometric ranges
for 2001$\,$QG$_{298}$ compare with the model predictions. Figure
\ref{Fig.Seven} overplots our 2010 data on the model lightcurves from Figure
\ref{Fig.Five}.  The first point to notice is that the new data are consistent
with and support the idea advanced by SJ04 that 2001$\,$QG$_{298}$ is a contact
binary (or at least a very elongated object), now being observed at a slightly
more pole-on geometry than in 2003.  Secondly, we find that the new data are
inconsistent with a low obliquity. In fact, the data points with smaller error
bars are best explained by an obliquity $\varepsilon=90\degr$, both in
rotational phase and photometric range (Figure \ref{Fig.Seven}).  We find the
lowest reduced $\chi^2$ value ($\tilde{\chi}^2=1.7$) for the
$\varepsilon=90\degr$ model.  Since a unit increase from the minimum reduced
$\chi^2$ roughly brackets a 1-$\sigma$ confidence interval
\citep{Press:1992:NRF:573178}, our data suggest an uncertainty in the obliquity
smaller than $\Delta\varepsilon=\pm30\degr$ (see Table\ \ref{Table.Four}).
This estimate neglects the effect of the uncertainty in the spin period and
rotational phase mentioned in Section 3.1.  To address this issue, we used a
Monte Carlo simulation that combines the photometric and timing uncertainties.
The goal is to quantify the probability that our data are best explained by a
given obliquity range.  We generated 1000 replicas of the 2010 lightcurve and
shifted each by a random phase drawn from a normal distribution with zero mean
and standard deviation 0.06 (see Section 3.1).  Then, each replicated data
point was nudged independently in the vertical (magnitude) direction by a
random amount taken from a normal distribution with zero mean and standard
deviation equal to its error bar. We found that 58\% of the artificial
lightcurves are best explained (in terms of $\chi^2$) by the model with
$\varepsilon=90\degr$.  These are followed by the $\varepsilon=75\degr$ model
with 12\%. We conclude that 70\% of the artificial lightcurves sampling the
uncertainty region are best explained by models with obliquities in the range
$\varepsilon=90\degr\pm30\degr$.  Finally, Figure \ref{Fig.Eight} shows that
for $\varepsilon=90\degr$ both surface types considered in LJ07 produce similar
results implying that the 2010 data are not adequate for discriminating between
the two. We conclude that the 2003 and 2010 lightcurves are simultaneously
explained by a single contact binary model with large obliquity,
$\varepsilon=90\degr\pm30\degr$. The simultaneous fit is shown in Figure
\ref{Fig.Ten}.

\section{Discussion}

The large obliquity of contact binary KBO 2001$\,$QG$_{298}$ is not unusual; it
is a property that is shared by other extreme shape objects in the solar
system.  Table\ \ref{Table.Five} lists the subset of bodies from the Planetary
Data System asteroid lightcurve database \citep{2009Icar..202..134Warner} with
maximum photometric variability larger than $\Delta m=0.9$ mag (indicative of
extreme shapes) and effective diameter larger than $D_\mathrm{e}=100$ km. We
have ignored smaller objects as they were probably more strongly affected by
collisions and non-gravitational effects that could interfere with their spins.
The bodies in Table\ \ref{Table.Five} are well studied and known to have
extreme shapes, and they all have obliquities in the range
$50\degr<\varepsilon<180\degr-50\degr$.  Although suggestive, the obliquities
of the objects in Table \ref{Table.Five} are nonetheless consistent with an
isotropic distribution. Randomly oriented spin poles fall in that same range of
obliquities with probability $p=\cos50\degr=0.64$. Hence, the chance that all
four objects in Table \ref{Table.Five} fall in that range, $p^4=0.17$, remains
high.  The sample size is too small to draw definite conclusions.

The origin of contact binaries remains unexplored. KBO binary formation
theories have focussed on the more easily identified resolved pairs
\citep{2002Icar..160..212W,2002Natur.420..643G,2004Natur.427..518Funato,2005MNRAS.360..401Astakhov,2008ApJ...673.1218Schlichting,2010AJ....140..785Nesvorny}.
The general idea behind wide-binary formation models is to identify a mechanism
that removes energy from a close but unbound pair of KBOs, converting it into a
gravitationally bound binary. Contact binaries are perhaps an end state of
initially ``hard'' (binding energy of the binary larger than the mean kinetic
energy of single objects) wide-binaries that have been progressively further
hardened by close encounters with passing, single bodies
\citep{1975MNRAS.173..729Heggie,2002Natur.420..643G}. Recently,
\citet{2009ApJ...699L..17Perets} proposed a contact binary formation mechanism
based on the combined effects of Kozai oscillations and tidal friction. 

Kozai oscillations \citep{1962AJ.....67..591Kozai,2007ApJ...669.1298Fabrycky}
occur in hierarchical triple systems (the Sun and the KBO binary in this case)
with large mutual orbital inclinations. In the solar system, the mutual
inclination between the inner and outer orbits of a binary is by definition the
obliquity, $\varepsilon$. If the initial obliquity $\varepsilon_0$ is larger
than the Kozai critical angle, typically
$\varepsilon_\mathrm{c}\approx40\degr$, the binary orbital eccentricity and
obliquity will oscillate in an anti-correlated fashion. As the obliquity falls
to a minimum of $\varepsilon=0\degr$ the eccentricity attains its maximum value
which depends on $\varepsilon_0$ and approaches unity for
$\varepsilon_0\approx90\degr$. Such large eccentricities will bring the binary
components very close (possibly even leading to collisions) allowing tidal
friction to dissipate energy and gradually shrink the binary semimajor axis.
\citet{2009ApJ...699L..17Perets} propose that repeated episodes of Kozai
oscillations and tidal friction could eventually lead to very compact systems
-- possibly contact binaries or highly elongated objects -- with eccentricities
$e\approx0$ and inclinations $40\degr<\varepsilon_0<180\degr-40\degr$.

\subsection{The fraction of contact binaries revisited}

Current estimates of the intrinsic fraction of contact binaries among the
Jupiter Trojans and the KBOs assume that their spins are isotropically oriented
\citep[LJ07;][]{2007AJ....134.1133Mann}. Since contact binaries are identified
from their high variability only if observed at an aspect angle sufficiently
close to $\theta=90\degr$, the apparent fraction must be corrected for the
probability of such favorable viewing geometry. LJ07 find minimum aspect angles
for positive contact binary identification of $\theta_\mathrm{lunar}=81\degr$
and $\theta_\mathrm{icy}=71\degr$ for lunar- and icy-type surfaces,
respectively.  The probability that a randomly oriented contact binary is
observed at aspect angle $\theta>\theta_\mathrm{min}$ is given by
$\cos\theta_\mathrm{min}$ \citep{2003Icar..161..174L} and amounts to
$p_\mathrm{lunar}=\cos81\degr=0.15$ and $p_\mathrm{icy}=\cos71\degr=0.33$ for
the extreme surface types considered.  Based on the discovery of
2001$\,$QG$_{298}$ in a sample of 34 KBOs (see SJ04), LJ07 calculate lower
limits to the fraction of contact binaries of $f_\mathrm{lunar}>20\%$ or
$f_\mathrm{icy}>9\%$ depending on the assumed scattering law. 

The estimated abundance will be higher if most contact binaries have high
obliquities. Figure\ \ref{Fig.Ten} illustrates the solid angle region available
to spin poles with high obliquities, in the range
$40\degr<\varepsilon<180\degr-40\degr$, and the fraction of that region
detectable from Earth. We numerically integrated the ``detectable'' and
``allowed'' solid angles and found their ratio (which equals the detection
probability) to be $p'_\mathrm{lunar}=0.11$ and $p'_\mathrm{icy}=0.24$ for the
two surface types. The lower detection probabilities with respect to the
isotropic case imply larger intrinsic abundances by $0.15/0.11=36\%$ (lunar
scattering) and $0.33/0.24=38\%$ (icy scattering). Therefore, if contact
binaries have obliquities in the range shown in Figure \ref{Fig.Ten} then the
lower limit abundance would increase to $f_\mathrm{lunar}>27\%$, assuming
lunar-scattering, and $f_\mathrm{icy}>12\%$, assuming icy scattering.

\section{Conclusions}

We have measured the visible lightcurve of Kuiper belt object
2001$\,$QG$_{298}$ and compared it to similar data collected in 2003. Our
findings can be summarised as follows:

\begin{enumerate}

\item The 2010 lightcurve of 2001$\,$QG$_{298}$ has a peak-to-peak range of
$\Delta m_{2010}=0.7\pm0.1$ mag, significantly lower than the photometric range
in 2003, $\Delta m_{2003}=1.14\pm0.04$ mag. The 2003 and 2010 lightcurves
appear in phase when shifted by an integer number of full rotations.

\item The change between the 2003 and 2010 lightcurves is most simply explained
if 2001$\,$QG$_{298}$ possesses a large obliquity,
$\varepsilon=90\degr\pm30\degr$. In that case, the lightcurve photometric range
should continue to decrease, reaching a minimum of $\Delta m\sim0.0-0.1$ mag in
2049.

\item Current estimates of the fraction of contact binaries in the Kuiper belt
assume that these objects have randomly oriented spins. If, as
2001$\,$QG$_{298}$, contact binaries tend to have large obliquities their
abundance may be larger than previously believed.

\end{enumerate}

\section*{Acknowledgments}

I thank David Jewitt for carefully reading and reviewing the manuscript. I also
thank Hilke Schlichting, Henry H. Hsieh and Maaike van Vlijmen for helpful
comments, and Alan Fitzsimmons for useful discussions. Hugo Lacerda Cruz and K. de Koek assisted in the preparation of the manuscript. I am very grateful for
the financial support from a Michael West Research Fellowship and from the
Royal Society in the form of a Newton Fellowship. I thank James McCormac for a
thorough tour of the Isaac Newton Telescope. The observations presented in this
paper were obtained as part of the Isaac Newton Group observing program
I/2010B/15.


\vfill
\eject
\clearpage
\FigOne

\vfill
\eject
\clearpage
\FigTwo

\vfill
\eject
\clearpage
\FigThree

\vfill
\eject
\clearpage
\FigFour

\vfill
\eject
\clearpage
\FigFive

\vfill
\eject
\clearpage
\FigSix

\vfill
\eject
\clearpage
\FigSeven

\vfill
\eject
\clearpage
\FigEight

\vfill
\eject
\clearpage
\FigNine

\vfill
\eject
\clearpage
\FigTen

\vfill
\eject
\clearpage
\input{tab1.tex}

\vfill
\eject
\clearpage
\input{tab2.tex}

\vfill
\eject
\clearpage
\input{tab3.tex}

\vfill
\eject
\clearpage
\input{tab4.tex}

\vfill
\eject
\clearpage
\input{tab5.tex}

\end{document}

%% file: tab1.tex
\begin{deluxetable}{ccccccccl}
  \tablecaption{Journal of Observations. \label{Table.One}}
   \tablewidth{0pt}
   \tablehead{
   \colhead{UT Date} & \colhead{$R$\tna} & \colhead{$\Delta$\tnb} & \colhead{$\alpha$\tnc} & \colhead{Filt.\tnd} & \colhead{Seeing\tne} & \colhead{Exp.\tnf} & \colhead{Notes} \\ 
   \colhead{} & \colhead{[AU]} & \colhead{[AU]} & \colhead{[\degr]} & \colhead{} & \colhead{[\arcsec]} & \colhead{[s]} & 
   }
   \startdata
2010 Aug 15 & 31.7618 & 31.1064 & 1.41 & $r$      & 0.9-1.5 & 600 & photometric \\
2010 Aug 16 & 31.7618 & 31.1053 & 1.41 & $r$      & 1.3-1.7 & 600 & clouds \\
2010 Aug 17 & 31.7618 & 31.1042 & 1.41 & $r$      & 0.9-1.3 & 600 & thin cirrus \\
   \enddata
  \tablecomments{All observations conducted at the 2.5-m Isaac Newton Telescope.}
  \tablenotetext{a}{Heliocentric distance in AU;}
  \tablenotetext{b}{Geocentric distance in AU;}
  \tablenotetext{c}{Phase angle in degrees;}
  \tablenotetext{d}{Filters used;}
  \tablenotetext{e}{Typical seeing in arcseconds;}
  \tablenotetext{f}{Typical integration time per frame in seconds.}
\end{deluxetable}

%% file: tab2.tex
\begin{deluxetable}{ccccccccl}
  \tablecaption{Calibration Stars. \label{Table.Two}}
   \tablewidth{0pt}
   \tablehead{
   \colhead{SDSS Name} & \colhead{$r$\tna} & \colhead{$g-r$\tnb}  \\ 
   \colhead{}          & \colhead{[mag]} & \colhead{[mag]}  
   }
   \startdata
SDSS J005028.45+020005.7 & $18.473 \pm 0.016$ & 1.39 \\ 
SDSS J005023.64+015726.9 & $18.951 \pm 0.025$ & 1.22 \\ 
SDSS J005026.24+015716.2 & $17.528 \pm 0.024$ & 0.56 \\ 
SDSS J005012.65+015638.6 & $19.131 \pm 0.026$ & 1.32 \\ 
SDSS J005011.82+015538.0 & $17.932 \pm 0.024$ & 0.97 \\
   \enddata
  \tablenotetext{a}{SDSS PSF $r$ magnitude;}
  \tablenotetext{b}{SDSS Model $g-r$ color;}
\end{deluxetable}

%% file: tab3.tex
\begin{deluxetable}{ccccccccccc}
   \tablecaption{2001$\,$QG$_{298}$ models. \label{Table.Three}}
   \tablewidth{0pt}
   \tablehead{
   \colhead{Surface\tna} & \colhead{$q$\tnb} & \colhead{$B/A$\tnc}& \colhead{$C/A$\tnc} & \colhead{$b/a$\tnd} & \colhead{$c/a$\tnd} & \colhead{$\omega^2/(\pi\,G\,\rho)$\tne} & \colhead{$d/(A+a)$\tnf} & \colhead{$\rho$\tng}
   }
   \startdata
     Lunar     &0.84&0.72&0.65&0.45&0.41&0.130&0.90&  0.59\\
     Icy       &0.44&0.85&0.77&0.53&0.49&0.116&1.09&  0.66\\
   \enddata
   \tablenotetext{a}{Surface scattering type;}
   \tablenotetext{b}{Mass ratio of the binary components;}
   \tablenotetext{c}{Axis ratios of primary;}
   \tablenotetext{d}{Axis ratios of the secondary;}
   \tablenotetext{e}{Dimensionless spin frequency of triaxial binary;}
   \tablenotetext{f}{Dimensionless binary orbital separation;}
   \tablenotetext{g}{Bulk density of the models (in g cm$^{-3}$).}
\end{deluxetable}

%% file: tab4.tex
\begin{deluxetable}{cccc}
  \tablecaption{Model Goodness of Fit. \label{Table.Four}}
   \tablewidth{0pt}
   \tablehead{
   \colhead{} & \multicolumn{2}{c}{Model Photometric Range\tnb} & \colhead{} \\ 
   \colhead{Obliquity\tna} & \colhead{$\Delta m_{03}$ [mag]} & \colhead{$\Delta m_{10}$ [mag]} & \colhead{$\chi^2_\mathrm{red}$\tnc}  
   }
   \startdata
    $\varepsilon=90\degr$ & 1.14 & 0.77 & \phn1.7 \\
    $\varepsilon=75\degr$ & 1.14 & 0.79 & \phn2.4 \\
    $\varepsilon=60\degr$ & 1.14 & 0.83 & \phn4.4 \\
    $\varepsilon=45\degr$ & 1.14 & 0.91 & \phn7.7 \\
    $\varepsilon=30\degr$ & 1.14 & 1.02 & 12.3 \\
    $\varepsilon=15\degr$ & 1.14 & 1.13 & 16.9 \\
    $\varepsilon=0\degr$\phn  & 1.14 & 1.14 & 18.8 \\
   \enddata
  \tablenotetext{a}{Obliquity of the model;}
  \tablenotetext{b}{Model photometric ranges in 2003 and 2010 assuming lunar-type scattering;}
  \tablenotetext{c}{Reduced $\chi^2$ value of 2010 model fit to the 2010 lightcurve.}
\end{deluxetable}

%% file: tab5.tex
\begin{deluxetable}{lccccccc}
  \tablecaption{Objects with Extreme Shapes. \label{Table.Five}}
   \tablewidth{0pt}
   \tablehead{
   \colhead{Name\tna} & \colhead{Group\tnb} & \colhead{$D_\mathrm{e}$\tnc} & \colhead{$P$\tnd} & \colhead{$\Delta m_\mathrm{max}$\tne} & \colhead{Obliquity\tnf} & \colhead{Shape\tng} & \colhead{Refs.} \\ 
   \colhead{} & \colhead{} & \colhead{[km]} & \colhead{[hr]} & \colhead{[mag]} & \colhead{[\degr]} & \colhead{} & \colhead{}
   }
   \startdata
     90 Antiope         & MBA & 120 &    16.5 & 0.90 & $53\pm2$ & Close binary & 1,2 \\ 
    216 Kleopatra       & MBA & 135 & \phn5.4 & 1.20 & $84\pm2$ & Bi-lobed & 3,4,5 \\ 
    624 Hektor          & Trojan & 230 & \phn6.9 & 1.10 & $98\pm5$ & Contact binary & 3,4,6,7 \\ 
 139775 2001 QG$_{298}$ & KBO & 250 &    13.8 & 1.14 & $90\pm30$ & Contact binary & 7,8,9\\
   \enddata
  \tablerefs{
  (1) \citet{2001A&A...378L..14M}; (2) \citet{2009Icar..203..102Descamps}; (3)
  \citet{1995P&SS...43..649D}; (4) \citet{2007Icar..192..223Kry}; (5)
  \citet{2011Icar..211.1022Descamps}; (6) \citet{1969AJ.....74..796D}; (7)
  \citet{2007AJ....133.1393L}; (8) \citet{2004AJ....127.3023S}; and (9) this
  work.}
  \tablenotetext{a}{Object designation;}
  \tablenotetext{b}{Object group (MBA=main-belt asteroid, Trojan=Jupiter Trojan, KBO=Kuiper belt object;}
  \tablenotetext{c}{Effective diameter;}
  \tablenotetext{d}{Approximate spin period;}
  \tablenotetext{e}{Maximum observed photometric range;}
  \tablenotetext{f}{Approximate obliquity.}
  \tablenotetext{g}{Most probable shape.}
\end{deluxetable}